\documentclass[twocolumn,fullpaper]{jpsj3}
\usepackage{graphicx}
\usepackage{wrapfig}
\usepackage{lastpage}
\usepackage[usenames]{color}
\usepackage{subfigure}

\title{Fractal Structure of Isothermal Lines and Loops \\
 on the Cosmic Microwave Background}

\author{Naoki \textsc{Kobayashi}\thanks{E-mail address: knaoki@phys.chuo-u.ac.jp}, Yoshihiro \textsc{Yamazaki}$^1$, Hiroto \textsc{Kuninaka}$^2$,
 Makoto \textsc{Katori},\\
 Mitsugu \textsc{Matsushita}, Satoki \textsc{Matsushita}$^3$ and Lung-Yih \textsc{Chiang}$^3$
}

\inst{
 Department of Physics, Chuo University, Kasuga, Bunkyo-ku, Tokyo, 112-8551\\
$^1$ Department of Physics, Waseda University, Ohkubo, Shinjuku, Tokyo 169-8555\\
$^2$ Faculty of Education, Mie University, Kurima-Machiya-cho, Tsu, Mie 514-8507\\
$^3$ Institute of Astronomy and Astrophysics, Academia Sinica, Taipei 10617, R.O.C.
}

\recdate{}

\abst{
The statistics of isothermal lines and loops of the Cosmic Microwave Background (CMB) 
 radiation on the sky map is studied and the fractal structure is confirmed in the radiation
 temperature fluctuation.
We estimate the fractal exponents, such as the fractal dimension
 $D_{\mathrm{e}}$ of the entire pattern of isothermal lines, the fractal dimension
 $D_{\mathrm{c}}$ of a single isothermal line, the exponent $\zeta$ in Kor\v{c}ak's law
 for the size distribution of isothermal loops, the two kind of Hurst exponents,
 $H_{\mathrm{e}}$ for the profile of the CMB radiation temperature, and $H_{\mathrm{c}}$
 for a single isothermal line.
We also perform fractal analysis of two artificial sky maps simulated by
 a standard model in physical cosmology, the WMAP best-fit $\Lambda$ Cold Dark Matter
 ($\Lambda$CDM) model, and by the Gaussian free model of rough surfaces.
The temperature fluctuations of the real CMB radiation and in the simulation using the $\Lambda$CDM
 model are non-Gaussian, in the sense that the displacement of isothermal lines and loops
 has an antipersistent property indicated by $H_{\mathrm{e}} \simeq 0.23 < 1/2$.}

\kword{%
cosmic microwave background radiation, isothermal lines and loops, fractal analysis, self-affinity, Hurst exponent
}

\begin{document}

\sloppy
\maketitle

\section{Introduction}

Our interesting feature of the Cosmic Microwave Background (CMB) radiation discovered
 by Penzias and Wilson \cite{PW65} is its uniformity over the full sky: the temperature
 of radiation is uniform at a value higher than one part per thousand.
This is now interpreted as evidence of the occurrence of the Big Bang and the rapid
 expansion of the universe during its very early stage.
The recent development of measurements using artificial satellites has revealed, however,
 the existence of small but definite temperature fluctuations, from which cosmologists are
 expecting to determine the origin of large-scale structures of galaxies and precise
 information for testing the Big Bang theory.\cite{PW99}
In the 1990s, the Cosmic Background Explorer (COBE) satellite conclusively detected
 a CMB radiation-temperature fluctuation in $\Delta T \sim 30 \mu\mathrm{K}$ on angular
 scales larger than $\sim 7^{\circ}$. \cite{Smoot92, Smoot07}
Following the launch of COBE, NASA constructed a second satellite, the Wilkinson Microwave Anisotropy
 Probe (WMAP), which was launched on 30 June 2001. 
WMAP has made a map of the temperature fluctuation of the CMB radiation with much
 higher resolution and accuracy than COBE. \cite{WMAP}
In the present paper, we discuss the temperature fluctuation of the CMB radiation from
 the viewpoint of statistical physics. 
In particular, after de Gouveia Dal Pino et al. analyzed the COBE data, \cite{GHHSVA95} we
 represent structures of fluctuations observed by the WMAP by
 producing a set of isothermal lines and loops on the sky map.
We extend the method of de Gouveia Dal Pino et al. by characterizing fractal structures,
 and in the present paper not only the fractal dimension of single isothermal line but also other
 fractal exponents of isothermal lines and loops are reported.


\begin{figure}
\centering
\includegraphics[width=1.0\linewidth]{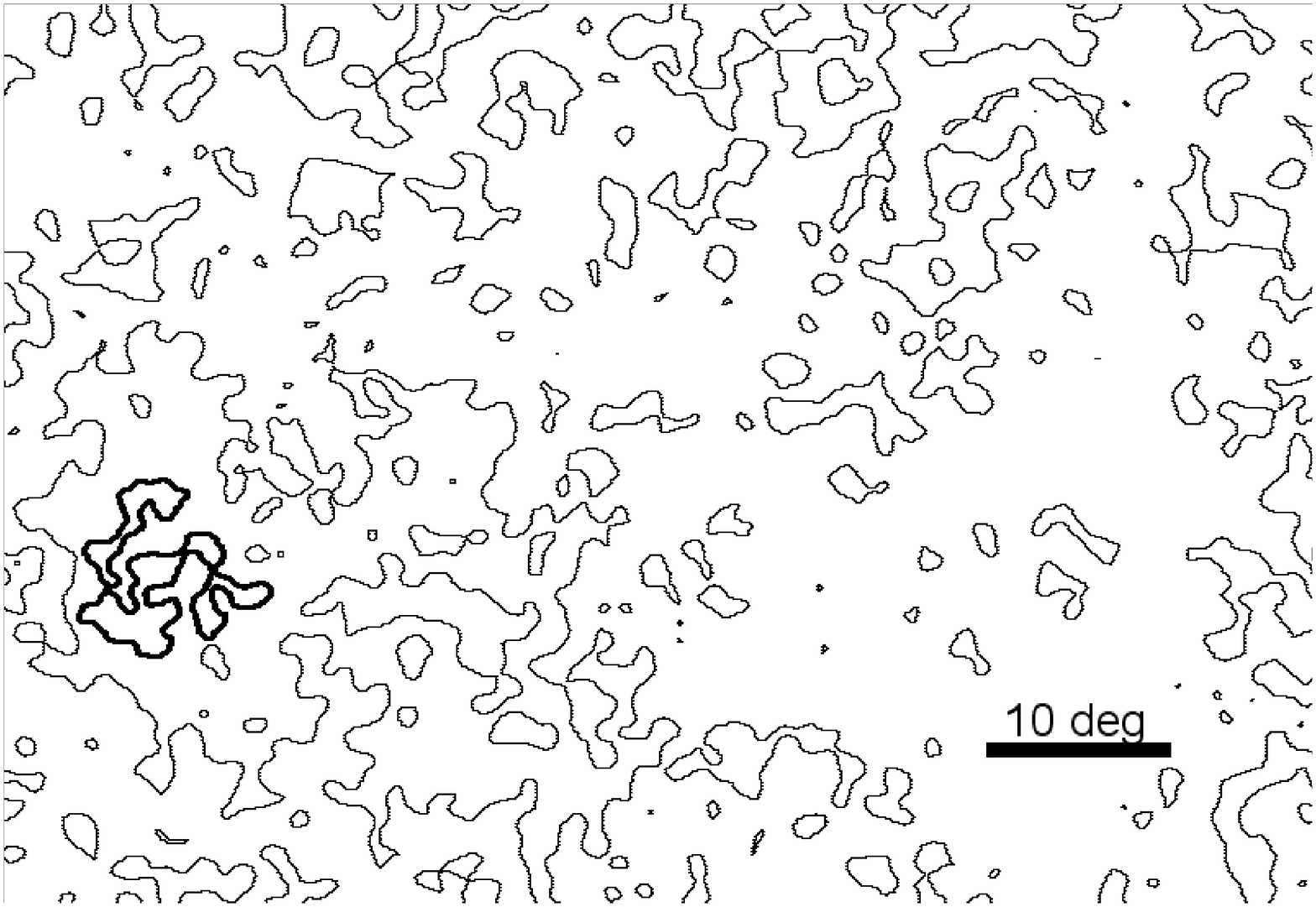}
\caption{Isothermal lines and loops of the CMB radiation on
 the sky map. Thick line represents one of the single isothermal lines.}
\label{fig:1}
\end{figure}

\section{Fractal analysis for isothermal lines and loops}

Figure 1 shows the set of isothermal lines and loops at an average temperature ($2.725$ K) in
 a partial region in the 5-year-CMB fluctuation measured by WMAP.
The isothermal lines are produced from data of the {\it WMAP Internal Linear Combination
(ILC) Map} provided at NASA's website. \cite{WMAP_SOS}
To characterize ``spatial" fluctuations of the CMB radiation-temperature shown in Fig. 1,
 we evaluate the fractal dimensions $D_{\mathrm{e}}$ for the entire set of isothermal lines
 and $D_{\mathrm{c}}$ for a single isothermal line (a contour, for example, a thick line shown
 in Fig. 1) on the CMB sky map.
Moreover, we study the cumulative size distribution of a set of isothermal loops
 larger than $x$ and show that it follows a power law with the characteristic exponent $\zeta$:
\begin{equation}
	N_{\mathrm{cum}} (x) = \int_{x}^{\infty} n(x') dx' \sim x^{- \zeta}.
\end{equation}
This power-law distribution is known as Kor\v{c}ak's law. \cite{Mandelbrot82, Feder88}

Now we report the results for the set of isothermal lines and loops at an average
 temperature.
We analyze the 5-year-CMB radiation corresponding to the ILC and foreground-reduced
 W-band maps, covering {\it ca.} $25^{\circ}$ on both sides of the line 
 that exactly divides the galactic east from the west in the 5-year-CMB sky maps measured by 
 WMAP, taken horizontally. \cite{WMAP_SOS}
The foreground reduced map was removed using the foreground template method of
 Hinshaw et al.\cite{Hinshaw_etal07} and Page et al.\cite{Page_etal07}.
Moreover, we use rotated maps in which the north pole is centered to avoid
 the high contamination of the data at the Galactic plane.
The fractal dimension $D_{\mathrm{e}}$ is determined using the box counting method.
We scale the number of boxes $N_{\mathrm{b}}(\epsilon)$, covering the
 entire set of isothermal lines, as a function of the box size $\epsilon$ as follows:
\begin{equation}
 N_{\mathrm{b}}(\epsilon) \sim \epsilon^{-D_{\mathrm{e}}}.
\end{equation}
Figure 2 shows the log-log plots of $N_{\mathrm{b}}(\epsilon)$ vs $\epsilon$ on the ILC map.
The slope in Fig. 2 yields the value of $D_{\mathrm{e}}$ as
\begin{equation}
 D_{\mathrm{e}} \simeq 1.77.
\end{equation}
On the other hand, the fractal dimension $D_{\mathrm{c}}$ is obtained by scaling
 the length $s(R_{\mathrm{g}})$ of an isothermal line (a contour) as a function of the
 radius of gyration
$R_{\mathrm{g}}^{2} = \sum_{i = 1}^{N_{\mathrm{T}}} (\mbox{\boldmath$r$}_\mathrm{i} - \mbox{\boldmath$r$}_\mathrm{c})^2 / N_{\mathrm{T}}$,
\begin{equation}
 s \sim R_{\mathrm{g}}^{D_{\mathrm{c}}},
\end{equation}
where $\mbox{\boldmath$r$}_\mathrm{c}$ is the center of mass of the isothermal line, $\mbox{\boldmath$r$}_\mathrm{i}$
 is the position of the $i$-th pixel of an isothermal line and $N_{\mathrm{T}}$ is the total number
 of pixels included in the line, respectively.
Figure 3 shows the log-log plots of $s$ versus $R_{\mathrm{g}}$ on the ILC map.
The slope in Fig. 3 yields $D_{\mathrm{c}}$ as
\begin{equation}
 D_{\mathrm{c}} \simeq 1.39.
\end{equation}
In the previous study, de Gouveia Dal Pino et al. obtained the fractal dimension
 $D_{\mathrm{c}} \simeq 1.42$ from the relation between
 the perimeter length and the area of an isothermal loop generated from the CMB sky map measured by COBE-DMR. \cite{GHHSVA95, BT02}
We also make a log-log plot of the cumulative size distribution of isothermal loops on the ILC map, as shown in Fig. 4.
Figure 4 shows that Kor\v{c}ak's law (1) is established and
\begin{equation}
 \zeta \simeq 0.87.
\end{equation}
The WMAP science working group reported that the ILC map provided a reliable estimate of
 the CMB signal on large angular scales greater than $\sim 10^{\circ}$. \cite{WMAP_suppl}
On the other hand, on smaller scales, they also reported that the ILC map might not be as
 reliable as it should be.
In order to check the validity of the fractal analysis for the ILC map, we reanalyzed the
 isothermal lines on the foreground-reduced W-band map.
The obtained values are $D_{\mathrm{e}} \simeq 1.74, D_{\mathrm{c}} \simeq 1.36$ and
 $\zeta \simeq 0.88$.
Because the differences are very small, we have concluded that the fractal analysis of the ILC map is reliable.


\begin{figure}
\centering
\includegraphics[width=0.9\linewidth]{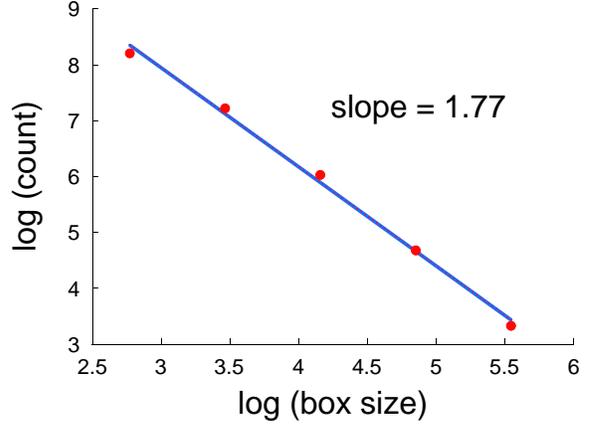}
\caption{Evaluation of $D_{\mathrm{e}}$ for the set of
 isothermal lines and loops on the CMB sky map.}
\label{fig:2}
\end{figure}


\begin{figure}
\centering
\includegraphics[width=1.0\linewidth]{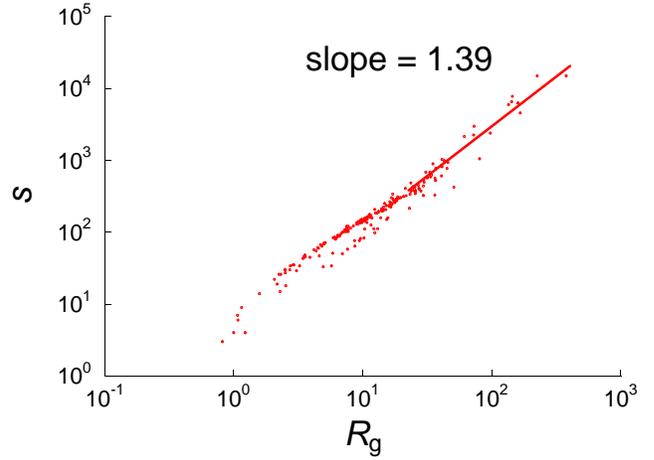}
\caption{Log-log plots of the length $s$ versus the radius of
 gyration $R_{\mathrm{g}}$ of a single isothermal line on the
 CMB sky map.}
\label{fig:3}
\end{figure}


\begin{figure}
\centering
\includegraphics[width=1.0\linewidth]{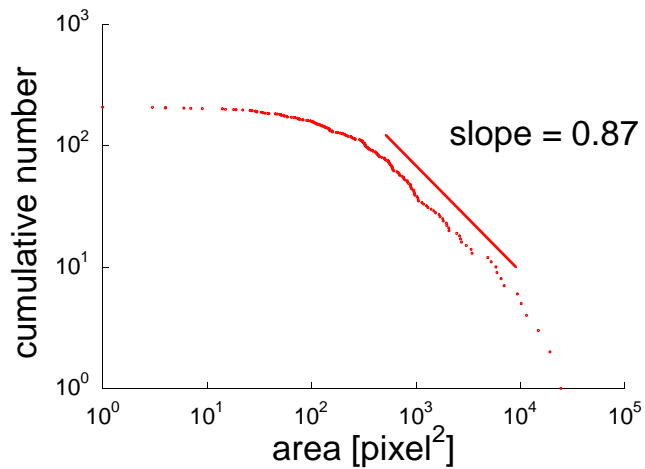}
\caption{Cumulative area distribution of isothermal loops
 on the CMB sky map.}
\label{fig:4}
\end{figure}


Rough surfaces, such as landforms, are characterized by the Hurst
 exponent. \cite{Mandelbrot82,  Feder88,Vicsek92, BS95}
The Hurst exponent $H$ is defined by the scaling exponent of the PDF $p_{H}(s, h)$
 of the fractional Brownian motion:
\begin{equation}
 p_{H}(s, h) = \frac{1}{\sqrt{2 \pi s^{2H}}} \exp \left\{ - \frac{h^2}{2 s^{2H}} \right\},
  \hspace{5mm} s > 0.
\end{equation}
For $H_{\mathrm{e}}$, we regard the ``time" $s$ as the horizontal position and the ``displacement" $h$
 as the altitude.
For $H_{\mathrm{c}}$, we regard $s$ as the length along the contour and $h$ as the fluctuation
 of a contour.
The characteristic exponents $D_{\mathrm{e}}, D_{\mathrm{c}}$, and $\zeta$ have scaling relations
 between them as reported by Matsushita et al.: \cite{MOH91}
\begin{eqnarray}
D_{\mathrm{e}} &=& 2 - H_{\mathrm{e}}, \\
D_{\mathrm{c}} &=& \frac{2}{1 + H_{\mathrm{c}}},  \\
\zeta &=& \frac{D_{\mathrm{e}}}{2} = \frac{2 - H_{\mathrm{e}}}{2},
\end{eqnarray}
where $H_{\mathrm{e}}$ is the Hurst exponent of the profile of a given rough surface and $H_{\mathrm{c}}$ is the Hurst
 exponent of a single isoheight line or loop of the surface.
In terms of the fractal dimension $D_{\mathrm{c}}$ of a single isothermal line, Kondev and Henley
 derived the scaling relation\cite{KH95}
\begin{equation}
 D_{\mathrm{c}} = \frac{3 - H_{\mathrm{e}}}{2}.
\end{equation}
These scaling relations have been confirmed by a variety of data analyses of real landforms and numerical
 simulations of rough surface models.
In general, when $H_{\mathrm{e}}= 1/2$, the displacements at different points are independent of
 each other (as in usual Brownian motion), and when $H_{\mathrm{e}} > 1/2$ (resp. $H_{\mathrm{e}} < 1/2$)
 they have a positive (resp. negative) correlation, which is called {\it persistence} (resp. {\it antipersistence}).
 \cite{MO89, MOH91, HTM92, KH95, KHS00, SKOM06}
The results for the CMB sky map are shown in the first column in Table I.
The Hurst exponents for the CMB sky map evaluated by the four relations eqs. (8)-(11)
 are all less than $1/2$.
That is, the displacement of the isothermal lines and loops of the CMB radiation has antipersistent property,
 which is ``noisier" than the usual Brownian motion.

\section{Numerical Simulations}


Chiang et al. performed a Monte Carlo simulation of the temperature
 fluctuations of the CMB sky map. \cite{CNVW03}
They used the standard model in physical cosmology, which is called the WMAP best-fit $\Lambda$ cold dark
matter ($\Lambda$CDM) model.
They have also provided an artificial sky map as reference for which the temperature fluctuation
 is generated from Gaussian white noise.
It may correspond to the limit $H_{\mathrm{e}} \to 0$ of the Gaussian rough surface model \cite{KH95, KHS00},
 which is called the Gaussian free field model in mathematical physics. \cite{SS09}
For these two artificial sky maps, we have also produced sets of isothermal lines and loops and
 performed the fractal analysis mentioned above.
The results are shown in the second and third columns in Table I.
The obtained fractal dimensions $D_{\mathrm{e}}$ and $D_{\mathrm{c}}$ 
 for the $\Lambda$CDM model are very close to those for the real CMB sky map.
In addition, Hurst exponents $H_{\mathrm{e}}$ and $H_{\mathrm{c}}$ obtained from the scaling relations
 eqs. (8)-(10) for the $\Lambda$CDM model is consistent with those of the CMB sky map.
On the other hand, the artificial sky map generated from the Gaussian white noise has $D_{\mathrm{e}} \simeq 2, D_{\mathrm{c}}
 \simeq 1.5 = 3/2, \zeta \simeq 1$, and $H_{\mathrm{e}} \simeq 0$.
These values are expected in the Gaussian free field model \cite{KH95, KHS00, SS09} and are definitely different from
 the actual CMB values except $\zeta$.


\begin{table}
\caption{Results of fractal analysis on the sets of isothermal
 lines. First column: for the real CMB sky map. Second column:
 for the artificial sky map simulated using the WMAP best-fit
 $\Lambda$CDM model. Third column: for the artificial sky map
 generated by the Gaussian white noise (Gaussian free model).
 Note that the mean values and standard deviation from 10 trials
 are shown for the artificial sky maps.}
\begin{tabular}{c|ccc}
\hline\hline
 & CMB map & $\Lambda$CDM model & white noise \\
\hline
$D_{\mathrm{e}}$ & 1.77 & 1.74 $\pm$ 0.03 & 1.90 $\pm$ 0.02 \\
$D_{\mathrm{c}}$ & 1.39 & 1.36 $\pm$ 0.07 & 1.53 $\pm$ 0.06 \\
$\zeta$ & 0.87 & 0.8 $\pm$ 0.1 & 0.9 $\pm$ 0.1 \\
$H_{\mathrm{e}}$ from (8) & 0.23 & 0.26 & 0.10 \\
$H_{\mathrm{e}}$ from (10) & 0.26 & 0.4 & 0.2 \\
$H_{\mathrm{e}}$ from (11) & 0.22 & 0.28 & -0.06 \\
$H_{\mathrm{c}}$ from (9) & 0.44 & 0.47 & 0.31 \\
\hline
\end{tabular}
\end{table}

\section{CMB and statistical physics}

One of the well-discussed topics regarding the recent progress in the conformal field theories and statistical
 physics of critical phenomena and random fractal patterns is the introduction of the Stochastic
 Loewner Evolution (SLE) by Schramm \cite{Schramm00}. 
SLE is a stochastic differential equation with a parameter, $\kappa$, generating a random
 fractal curve on a plane.
In particular, SLE with some characteristic values of $\kappa$ describes critical curves
 in statistical physics such as the percolation exploration process and the phase boundary
 of the critical Ising model, as well as the continuum limits of the loop-erased random walk,
 the self-avoiding random walk (SAW), the random Peano curve, and other fractal curves. \cite{KN04, Gruzberg06, SRS08, SDR10}
The set of isothermal lines generated from the Gaussian white noise has
 $D_{\mathrm{c}} \simeq 1.50 = 3/2$ (the Gaussian free field model) as previously mentioned.
This correponds to the SLE curve with $\kappa = 4$. \cite{SS09}
We note that the fractal exponents $D_{\mathrm{c}} \simeq 1.39$ and $H_{\mathrm{e}} \simeq
 0.23$, evaluated in the present study for the set of isothermal lines and loops on the CMB sky map,
 can be compared with the exponents of the phase boundary of the critical Ising model,
 $D_{\mathrm{c}} = 11/8 = 1.375$ and $H_{\mathrm{e}} = 1/4$. This corresponds to
 the SLE curve with $\kappa = 3$. \cite{S09}
Note that the SAW corresponding to the SLE$_{8/3}$ curve
 has the fractal exponents $D_{\mathrm{c}} = 4/3 = 1.\dot{3}$ and $H_{\mathrm{e}} = 1/3 = 0.\dot{3}$.

In general, the Hurst exponent on the $(s, h)$-plane is given by the self-affine exponents
 $\nu_s$ and $\nu_h$ in each direction: \cite{MO89}
\begin{equation}
 H = \frac{\nu_h}{\nu_s}.
\end{equation}
The exponents $\nu_s$ and $\nu_h$ are obtained using the asymptotics $s \sim N^{\nu_s}$ and
 $h \sim N^{\nu_h}$, where $s$ and $h$ are the root-square-mean displacements of the $s$ and $h$ components
 of the position of each pixel in a curve, respectively,  and $N$ is the curve length
 between two arbitrary points on the pattern.
Formula (12) makes it possible to evaluate the Hurst exponents directly from the
 profile of the temperature fluctuation and each isothermal loop on the CMB sky map.
Moreover, the measurement of the self-affine exponents will enable us to discuss the
 validity of the scaling relations and the connection to other fractal interface
 models such as the Karder-Parisi-Zhang (KPZ) equation. \cite{KPZ86, BS95}
Multifractal analysis is also useful in characterizing the distribution of nonuniform fluctuations. \cite{MS87}
In a previous study, Diego et. al. have applied multifractal analysis to the study of the CMB
 temperature fluctuation measured by COBE-DMR to calculate the generalized dimension $D_q$
 and $f(\alpha)$ spectrum. \cite{DMSMM99}
The CMB temperature fluctuation data measured by WMAP and PLANCK \cite{PLANCK} satellites should be
 analysed using the multifractal method.
The set of isothermal loops on the CMB sky map has the property showing that the size of a loop
 changes significantly owing to the change in the threshold temperature to draw isothermal loops.
The relation between a threshold temperature and the size of a loop is similar to the
 relation between occupation probability and the size of a percolation cluster. \cite{SA94}
This similarity raises a ``percolation problem on the CMB sky map".
These studies are currently under investigation by our collaborators.

\section{Summary}

We have investigated the fractal exponents $D_{\mathrm{e}}$ of the entire
 set of isothermal lines, $D_{\mathrm{c}}$ of each single isothermal line, and $\zeta$
 for Kor\v{c}ak's law of the CMB radiation-temperature fluctuations measured by WMAP satellite.
Moreover, we have determined the Hurst exponents $H_{\mathrm{e}}$ and $H_{\mathrm{c}}$
 that respectively characterize the self-affinity of the radiation-temperature profile and each isothermal line
 on the galactic coordinates obtainded by applying the scaling relations eqs. (8)-(11).
In the same way, we have also obtained the fractal exponents of two artificial sky maps generated using the WMAP
 best-fit $\Lambda$CDM model and the Gaussian free model of rough
 surfaces. \cite{CNVW03}
The values of the various scaling exponents shown in Table I imply that 1) the fractal
 exponents $D_{\mathrm{e}}, D_{\mathrm{c}}, \zeta, H_{\mathrm{e}}$, and $H_{\mathrm{c}}$ of the real CMB sky map and
 those of the $\Lambda$CDM model coincide with each other,
 and 2) the evaluated $H_{\mathrm{e}}$ for the CMB radiation is less than $1/2$, that is, the temperature
 fluctuation has an antipersistent property.
In cosmology, it is important to check whether the primordial fluctuation 
 in the universe is Gaussian.
To check for Gaussianity, the statistical properties of the CMB radiation-temperature
 fluctuation have been investigated by various methods, such as the use of angular
 bispectrum \cite{K_etal03, YW08, K_etal09, SSZ09, HKMTY09}
 Minkowski functionals \cite{Schmalzing_Gorski98, K_etal03, K_etal09, HKMTY09},
 phase mapping techniques \cite{CCN02}, and fractal analysis \cite{DMSMM99, MMDMSP99}.
On the basis of these analyses, it was found that the observations of the CMB radiation-temperature
 fluctuation are compatible with the Gaussian hypothesis. \cite{K_etal03, K_etal09, HKMTY09,
 CCN02, DMSMM99, BKMR04}
We note, however, that Yadov and Wandelt presented evidence of the primordial non-Gaussianity
 $f_{\mathrm{NL}} \neq 0$, where $f_{\mathrm{NL}}$ denotes the nonlinearity parameter of the local
 type. \cite{YW08}

In the present study, we confirmed that the ensemble of isothermal lines and loops of the CMB has
 the fractal property and that the methods for analyzing random surfaces and their isoheight lines
 and loops developed in the statistical mechanics and fractal physics are very useful.
Here, we have reported that the fluctuation of isothermal lines and loops of the CMB
 radiation has antipersistent property, since the Hurst exponent $H_{\rm e} < 1/2$.
We think that our present study will open a new field, where cosmologists and
 statistical physicists can enjoy communications with each other to establish better understanding
 of the universe.
\begin{acknowledgements}
The present authors are grateful for the use of the Legacy Archive for Microwave
 Background Data Analysis (LAMBDA) by NASA.
N. K.  would like to thank J. Wakita, R. Nakahira, M. Higuchi, Y. Nishikiori
 and S. Andraus for many stimulating discussions and he is supported in part by
 Grant-in-Aid for Young Scientists (B) (No. 22700739) of Japan Society for the Promotion of Science.
M. K. is supported in part by Grant-in-Aid for Scientific Research (No. 21540397)
 of Japan Society for the Promotion of Science.
M. M. is supported in part by Grant-in-Aid for Scientific Research (No. 22540399)
 of Japan Society for the Promotion of Science.
\end{acknowledgements}


\end{document}